\documentclass[11pt,twoside]{article}
\usepackage{asp2006}
\usepackage{psfig}
\usepackage{epsf}
\usepackage{graphics}
\usepackage{lscape}
\def\edcomment#1{\iffalse\marginpar{\raggedright\sl#1\/}\else\relax\fi}
\reversemarginpar

\begin{document}
\title{Cool Stars in Hot Places}
\author{S. T. Megeath}
\affil{Ritter Observatory, Department of Physics and Astronomy, University of Toledo, 
Toledo, OH 43606}
\author{E. Gaidos}
\affil{Department of Geology \& Geophysics, University of Hawaii, Honolulu, HI 96822}
\author{J. J. Hester}
\affil{Arizona State University, Department of Physics \& Astronomy, Tempe, AZ 85287}
\author{F. C. Adams}
\affil{Physics Department, University of Michigan, Ann Arbor, MI 48109}
\author{J. Bally}
\affil{Center for Astrophysics and Space Astronomy, University of Colorado, Boulder, CO 80309}
\author{J.-E. Lee}
\affil{Physics and Astronomy Department, The University of California at Los Angeles, Los Angeles, CA 90095}
\author{S. Wolk}
\affil{Harvard Smithsonian Center for Astrophysics,  Cambridge, MA 02138}

\begin{abstract}

During the last three decades, evidence has mounted that star and
planet formation is not an isolated process, but is influenced by
current and previous generations of stars.  Although cool stars form
in a range of environments, from isolated globules to rich embedded
clusters, the influences of other stars on cool star and planet
formation may be most significant in embedded clusters, where hundreds
to thousands of cool stars form in close proximity to OB stars.  At
the cool stars 14 meeting, a splinter session was convened to discuss
the role of environment in the formation of cool stars and planetary
systems; with an emphasis on the ``hot'' environment found in rich
clusters.  We review here the basic results, ideas and questions
presented at the session. We have organized this contribution into
five basic questions: what is the typical environment of cool star
formation, what role do hot star play in cool star formation, what
role does environment play in planet formation, what is the role of
hot star winds and supernovae, and what was the formation environment
of the Sun?  The intention is to review progress made in addressing
each question, and to underscore areas of agreement and contention.

\end{abstract}

\vspace{-0.5cm}

\section{What is the Typical Environment of Cool Star Formation?}

Cool stars form in a range of environments, from isolated Bok
globules, to modest sized clusters containing 100-200 stars, and
finally to large, dense clusters with thousands of cool stars and several
to tens of OB stars. This is in sharp contrast to OB stars, which form
almost entirely in large clusters. This motivates the question: in
what environment do most cool stars form?

Surveys of the molecular gas in our Galaxy indicate that most of the
cold molecular gas is in giant molecular clouds (GMCs) with masses of
$10^5$ to $10^6$~M$_{\odot}$ \citep{heyer1998}. These massive
molecular clouds are thought to form entire associations of hot OB
stars as well thousands of low mass stars. Coupled with analyses
indicating that 80-90\% of cool stars form in large clusters
\citep{porras2003,lada2003,carp2000}; these results seemed to point to
a galaxy in which the vast majority of cool star formation takes place in
rich crowded clusters in close proximity to hot stars.  However, since
there was little information on the numbers of isolated stars, the
analyses of \citet{porras2003} and \citet{lada2003} considered only
stars in groups and clusters. In an analysis of the 2MASS point source
catalog toward several molecular clouds, \citet{carp2000} found
evidence for substantial numbers of isolated stars, but the estimates
contained significant uncertainties.

More recently, surveys of giant molecular clouds with the {\it
Spitzer} space telescope provided the means to identify isolated young
stars and protostars through the infrared excesses from their disks
and envelopes \citep{allen2007}.  {\it Spitzer} surveys of four giant
molecular clouds containing young massive hot stars, the Orion~A
cloud, Orion~B cloud, Cep~OB3 cloud and Mon~R2 cloud, show that in
addition to clusters associated with regions of massive star
formation, there are large number of stars in small groups or
isolation.  In these clouds, 46\% of the young stars with
excesses are found in clusters with over 90 sources, 11\% are found in
small clusters of 90-30 stars, 8\% in groups of 30-10 stars, and the
remaining 35\% in groups with less than 10 members or isolation
\citep{megeath2007,gutermuth2007}.  About 33\% of the stars are found
in the two largest clusters with over 700 members each. Thus, although
most cool stars may form in OB associations, young cool stars in OB
associations are not found primarily in large clusters.  Instead, they
are found in a range of environments, with a significant fraction of
stars forming in relative isolation several to tens of parsecs away
from the nearest OB stars (Fig.~1).

\begin{figure}[!ht]
\begin{center}
\plotone{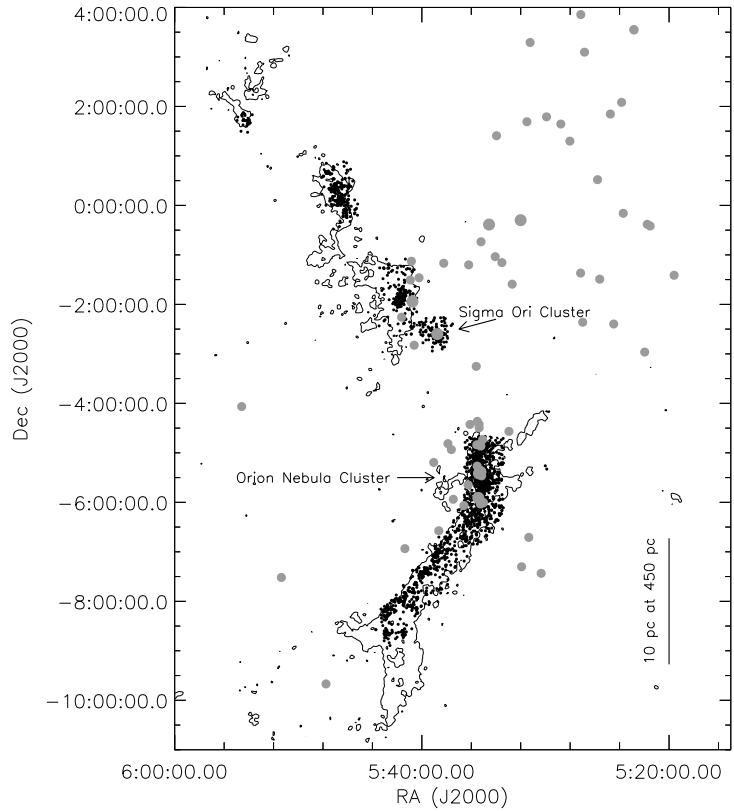}
\vskip 0.15 in
\caption{The Orion OB association.  The contours show an A$_V$ map of
the Orion region made from the 2MASS database (Gutermuth, p. com.), the
gray circles are the O stars (large circles) and B stars (small circles)
from Brown et al. (1994) and the dots are {\it Spitzer} identified
young cool stars and protostars (Megeath et al. 2007; Hernandez et
al. 2007). Only regions of high molecular column density and the
$\sigma$~Ori cluster have been surveyed by Spitzer, and many more
young cool stars certainly exist in the OB association}.
\end{center}
\end{figure}

\section{What Role do Hot Stars play in Cool Star Formation?}

Although cool stars dominate star-forming regions in both number and
total stellar mass, hot stars are thought to be the primary agents of
molecular cloud evolution.  The extreme-UV radiation from young O and
early B-type stars photoionizes the surfaces of molecular clouds,
resulting in flows of ionized gas which erode the clouds.  Far-UV
radiation may play a similar role in regions where only B-stars are
present by heating and photodissociating the molecular gas. Clusters
with O and/or B stars and ages of only a few million years appear to
have partially or fully dispersed their molecular clouds. An example
is the 2.5~Myr old $\sigma$~Ori cluster \citep{sherry2004}; this
cluster sits {\it outside} the Orion~B cloud (Fig.~1).  Other examples
are discussed in \citet{allen2007}.  It is estimated that only 10\% of
embedded clusters survive gas dispersal and presist as clusters for
more than 10~Myr \citep{lada2003}.

The detection of Evaporating Gaseous Globules (EGGs), 1000 AU diameter
photoevaporating dark globules, demonstrated that hot stars may
directly impact protostellar evolution \citep{hester1996}. EGGs appear
to be protostellar or prestellar cores which emerge from their
parental clouds as the surrounding lower density gas is ionized
\citep{hester2005}.  In M16, 15\% of the EGGs contain embedded stars,
indicating that they are the sites of recent or ongoing star formation
\citep{mccaughrean2002}.  This suggests that hot stars can directly
affect protostellar evolution by photoevaporating the infalling gas
and limiting the ultimate mass of the nascent star.  However, it is
not known what fraction of stars emerge from their clouds in EGGs.

A more controversial issue is whether OB stars trigger cool star
formation.  This possibility has been discussed in the literature for
decades \citep[e.g.][]{elmegreen1977}.  \citet{hester2005} proposed
that in regions with hot stars, cool star formation is driven {\it
primarily} by shock fronts preceeding advancing ionization fronts. The
shock fronts overtake and compress pre-existing density enhancements,
inducing collapse and the formation of clusters of low mass stars.
Evidence for this is found in the detection of clusters of young stars
at the surfaces of molecular faces being eroded by hot stars
\citep{sugitani1995,megeath2004,allen2005}. However, additional evidence, such as
the detection of the shock fronts, is needed to determine whether the
clusters have been triggered, or whether they are regions of
ongoing star formation which have been overtaken by ionization
fronts \citep{megeath1997}.

Although there is growing evidence that triggering does happen, it is
not clear what fraction of cool star formation is triggered.
Assessing the overall importance of triggered star formation can be
difficult due to the rapid evolution and even rapid motions of OB
stars.  For example, \citet{hoogerwerf2001} argued that the
interaction of the $\iota$ Ori binary system with a second system led
to the ejection of the runaway stars AE Aur and $\mu$~Col 2.5~Myr ago
(both are 09.5 stars).  Although they suggested that these stars
originated in the Orion Nebula Cluster, the lack of a visible HII
region surrounding $\iota$ Or, an O9~III star which in projection
appears conicident with the Orion~A cloud, suggests that it is several
to tens of parsecs away from the Orion A molecular cloud and is part
of the 5~Myr OB1c association \citep{brown1994}. At the time of their
ejection, these three O--stars may have had a significant impact
on the Orion~A cloud, and could have been responsible for triggering
star formation in the Orion Nebula Cluster.  Another possible example
is the LDN~1551 dark cloud in the Taurus dark cloud complex.  This
cloud has a cometary morphology with the ``head'' of the comet pointing
toward the Orion constellation.  \citet{moriarty2006} argued that the
cometary shape may be due to the interaction of LDN~1551 (149 pc from
the Sun) with the B8I star Rigel (Hipparcos distance is 240 pc) and the
M2I star Betelgeuse (Hipparcos distance 130 pc).  The high proper
motion of Betelgeuse would place it southeast of LDN~1551 several million
years ago; hence, both Betelgeuse and Rigel could have plausibly
interacted with LDN~1551, creating the cometary morphology.

These observations demonstrate the difficulties in determining causal
relationships between subsequent generations of star formation and establishing
the importance of triggering.
Although ongoing triggering can be identified by the 
detection of clusters near ionization fronts, in many cases, evidence
of triggering may be erased by the evolution and motion of massive
stars.  

\section{What Role does Environment Play in Planet Formation?}

Environment may also play a role in planet formation by altering the
properties of protoplanetary disks.  We discuss here two mechanisms:
tidal interactions between stars in clusters and the photo-ablation of
disks by UV photons from nearby OB stars.

Tidal interactions occur when a disk around a star in a cluster is
distorted or stripped during a close encounter with another cluster
member. Such interactions appear to be unimportant.  Adopting a
stellar density of $10^4$~pc$^{-3}$ (the {\it peak} density for many
embedded clusters) and assuming virialized velocities,
\citet{gutermuth2005} used a simple mean free path argument to
estimate the frequency of close approaches.  They estimated that even
in the dense, central cores of clusters, close approaches at distances
of 100~AU would occur once in a 10~Myr interval. However, the high
stellar densities assumed by Gutermuth et al. may only persist for a
few million years before the clusters begin to expand. This result is
supported by N-body simulations of bound clusters which show that such
interactions are rare over the lifetime of an embedded cluster
\citep{adams2006,throop2007}.  \citet{adams2006} find that each star
in a 1000-member (initially) embedded cluster will experience one
close-approach within 700-4000~AU over a 10~Myr interval.  This
distance is more than three times the typical radius of observed
circumstellar disks in nearby dark clouds \citep{andrews2006} and
much larger than the size of the Solar System.  Since the adopted
timescale for gas removal in these simulations was 5~Myr, longer than
the observed timescale (Sec.~2), the close-approach distances should
be considered lower limits.  In summary, the results from three
independent investigations are in agreement; unless embedded clusters
exist in our galaxy with much higher stellar densities than observed
in nearby regions such as the Orion Nebula Cluster, tidal interactions
in clusters rarely influence disk evolution and planet formation.

In contrast, photoevaporation of disks by nearby OB stars appears to
be a much more influential process.  The UV radiation from the OB
stars heats the gas in disks through photoionization and
photodissociation, resulting in flows of gas off the disks.  This
process was discovered in VLA and HST observations of young stars in
the Orion Nebula \citep{churchwell1987,odell1994}. The inferred mass
loss rates were $10^{-7}$~M$_{\odot}$, suggesting disk lifetimes of
only a few hundred thousand years \citep{bally1998}.  However, the
mass loss occurs in the outer disk where the thermal velocity of
ionized gas exceeds the escape velocity from the star, and the gas in
the inner disk may not be strongly affected.  More recent calculations
include the effect of the far-UV radiation and the time dependent
nature of the UV-field as the stars orbit within the cluster
potential.  \citet{adams2004} calculate the mass loss from a disk as a
function of the intensity of the far-UV radiation field. They find the
radiation field can truncate a disk to the size of our solar system in
several million years; the exact radius depends on the duration of the
exposure to UV radiation, the intensity of the UV radiation, and the
mass of the central star \citep{adams2004}. \citet{throop2007} use
N-body simulations to calculate the time dependent flux of UV
radiation incident on a young star with disk as it orbits in a cluster
which contains OB stars in its center. They find that typical stars
experience only a brief exposure to intense UV as they pass within
10,000~AU of the central OB stars. Consequently, the UV flux incident
on a disk varies in an stochastic manner over the lifetime of the
cluster.

Recently, \citet{throop2005} proposed that the photoevaporation of
disks may in fact trigger the formation of planets.  In their model,
grain growth and dust settling concentrates dust grains in the
midplane of the disk. Consequently, the ablation of the gas from the
disk surface (as well as the remaining dust grains entrained in the
gas) reduces the ratio of the gas surface density to dust surface
density.  If the surface density of gas is reduced to less then 10
times the dust density, the disk becomes unstable to gravitational
collapse \citep{sekiya1998,youdin2002}.

Although photoevaporation may be important in rich embedded clusters
with OB stars, many young cool stars in OB associations are not found
in such clusters.  Young cool stars with disks identified in the
Spitzer survey of the Orion A cloud have a median projected distance
of 4.1 ~pc to the nearest O to B0 star, and a median projected distance
of 2.1 pc to the nearest B1-B3 stars \citep{megeath2007}.  Hence, in
OB associations, most cool stars may form at large distances from the
central OB stars and are unaffected by their UV radiation.

\section{What is Role of Hot Star Winds and Supernovae?}

Chandra X-ray observations of young stellar clusters have detected
diffuse X-ray emission in nine regions.  The total luminosities of
this gas range from $1 - 200 \times 10^{33}$~erg~s$^{-1}$
\citep{wolk2002,townsley2006}.  Although supernovae could generate
this gas, in most cases the diffuse gas appears to be generated by
stellar winds from massive stars colliding with other winds or the
surrounding HII region.  However, in the Carina region, a component of
hot gas enriched in Fe was likely created by a supernova
\citep{townsley2006}. The impact of the extremely hot gas on star and
planet formation is not well understood. In addition to destroying the
surrounding the cloud, the blast waves from a supernova could compress
surrounding cores of gas causing them to collapse into stars
\citep{boss1995,melioli2006}.  Disks can survive at distances of $\le
1$~pc from a supernova \citep{chevalier2000}; however, these disks
will be heated by the radiation and blast wave, and may also be
stripped by the blast wave when the disks are only 0.25~pc from the
supernova \citep{chevalier2000}.  The hot X-ray gas created by winds
may fill bubbles within the larger HII region.  This hot, low density
gas would be transparent to UV photons, and hence any young
stars within the bubble may be exposed to a more intense UV field than
those in the surrounding HII region.

\section{What Was the Formation Environment of the Sun?}

Did our Sun also form in the ``hot'' environment of a large embedded
cluster?  \citet{tremaine1991} and \citet{gaidos1995} proposed that
our Solar System might preserve dynamical evidence of its birth
environment.  \citet{gaidos1995} and \citet{adams2001} used the low
inclination and eccentricity of Neptune to place constraints on the
time-integrated tidal field of a cluster and the closest stellar
passage.  However, such reasoning must now be re-examined in light of
the expectation that most embedded clusters expand and disperse in a
few Myr (although some clusters would form bound open clusters,
Sec.~2) and the realization that Neptune (and Uranus) migrated outward
to its present orbit by scattering in a residual planetesimal disk, a
process that was probably not completed until after a parental cluster 
dispersed \citep{hahn2005}.  Scattering inside the disk
itself, which dampens any non-circular motion, could have produced the
low eccentricity and inclination observed today.  Similar arguments
can be made that other parts of the outer Solar System (the
Edgewood-Kuiper belt, Oort Cloud) formed after the cluster evaporated
\citep{levison2003}. \citet{kenyon2004} and \citet{morbidelli2004}
proposed that Sedna, a member of the scattered Kuiper Belt, was
produced by the close passage of a star, but there are other
explanations \citep{barucci2005,gladman2006}.  Thus, it is likely that
the structure of the outer Solar System post-dates an embedded cluster
phase.

The strongest evidence for an early cluster environment is the
inferred presence of short-lived radionuclides (SLRs) during the
formation of solids now found in meteorites.  There are at least three
possible sources of SLRs: particle irradiation within the primordial
solar nebula, the wind from a nearby AGB star, and the wind and/or
supernova ejecta from a nearby massive star.  The discovery of
$^{60}$Fe in the early Solar System \citep{Tachibana2003} firmly
establishes that the Sun formed in a rich cluster containing massive
stars \citep{hester2004,hester2005}.  Neutron-rich isotopes such as
$^{60}$Fe cannot be produced by particle irradiation. The 
uniform distribution of the SLR $^{26}Al$ makes it unlikely it
was produced by irradiation \citep{thrane2006}. Finally, it is
statistically unlikely that the SLRs originated in an AGB star
\citep{kastner1994}.

Further evidence is found in the mass-independent fractionation of the
oxygen isotopes ($^{17}$O and $^{18}$O) in meteorites.  Following a
proposal by \citet{clayton1973}, \citet{lee2007} have made a
theoretical analysis of the time-dependent chemistry in a collapsing
envelope subjected to an external UV field.  Due to ``self-shielding''
of the much more abundant C$^{16}$O, the UV field preferentially
dissociates C$^{18}$O and C$^{17}$O, producing an enhancement of
$^{18}$O and $^{17}$O in the gaseous envelope.  These heavier isotopes
are then incorporated (as water) into ice grains and transported into
the inner region of the solar nebula.  This process depends on the
intensity of the external UV radiation field (from OB stars) so that
the measured fractionation can constrain the formation environment of
the Sun. \citet{lee2007} conclude that the observed isotopic ratios
are best explained by a radiation field 10$^5$ greater than the
interstellar field, again supporting the presence of nearby massive
stars.

The current evidence firmly indicates that the Sun formed in a hot
environment enriched by the ejecta of one or more nearby supernova;
however, there is a continuing debate over how the solar nebula was
enriched.  \citet{cameron1995} argued that the enrichment occurred
when the collapse of the proto-solar molecular cloud was triggered by
the blast wave of a supernova \citep[also
see][]{vanhala2002}. \citet{hester2005} question whether this process
could enrich the collapsing molecular gas.  Alternatively,
the protostellar envelope of the Sun may have been directly enriched
while collapsing onto the proto-Sun \citep{looney2006}.  For example,
if the solar system formed in an EGG, then it may have been subjected
to  a blast wave from a supernova.  Finally, the SLRs may have been
injected directly into the disk of the solar nebula when the Sun was
in its T-Tauri phase; a possible mechanism for this is the ``aerogel''
model, in which grains in SN ejecta are deaccelerated and vaporized
within the gaseous primordial disk \citep{ouellette2005}.  This
scenario is supported by observations showing that 40\% of disks may
persist for 4~Myr \citep{hernandez2007}, the lifetime of a
60~M$_{\odot}$ star.

Recent quantitative analyses have constrained the distance between the
Sun and the supernova from which the SLRs presumably originated. If
the enrichment occurred while the Sun was in a T~Tauri phase with a
200~AU disk \citep{andrews2006}, the estimated distance is between
0.04-0.4~pc \citep{looney2006,elinger2007}.  If the enrichment
occurred in the protostellar phase (5000 AU diameter), the estimated
distance is between 0.12- 1.6~pc \citep{looney2006}.  The question has
been raised whether these distances are consistent with observations
showing that embedded clusters largely disrupt their parental cloud
and disperse in a few million years (see Sec.~2).  The dispersal of
the molecular gas makes the presence of nearby protostars unlikely,
and the subsequent expansion of the cluster make the presence of young
stars with disks less likely.  There are possible solutions to this
problem. The Sun may have remained close to a hot star as the cluster
dispersed. Only one low mass star with a disk is found within
a projected distance of $\sim 0.3$~pc of the O6 star HD206267 in the
4~Myr old IC 1396 association \citep{sicilia2006}, suggesting that
this may be rare occurrence.  The Sun may have been a bound companion
to a massive star, such as the companions with disks found around the
OB stars comprising the Orion Trapezium \citep{schertl2003}; however,
it unclear how long such a disk may survive. The Sun could have formed
in a massive embedded cluster which evolved into a bound open
cluster.  In this case, the solar system would have to survive
photoevaporation and perturbations from tidal interactions as it
orbited within the cluster \citep{adams2001}.  Finally, the solar
system may have been enriched by the combined ejecta of many supernova
\citep{hester2005, williams2007}. Additional data on SLRs in
meteorites, detailed modeling of the evolution and dispersal of
embedded clusters, and the study of other planetary systems in hot
environments should bring a more detailed understanding
of our Sun's formation environment.

The presence of SLRs may have had a significant impact on planet
formation in the solar nebula.  Radioactive decay of $^{26}$Al and
$^{60}$Fe provides by far the largest source of energy for melting and
differentiating planetesimals in the early Solar System
\citep{bizzarro2005,hevey2006}.  In summary, it has been amply
demonstrated by observation and theory that environment plays a
significant role in the formation of cool stars and planets.  A
comprehensive understanding of star and planet formation must not
treat young stars and protoplanetary solely as isolated objects, but
as parts of larger associations and clusters in which the formation of
cool and hot stars are inextricably linked.












\end{document}